# A Bespoke Forensics GIS Tool

Almar Tillekens

The Netherland
almar@digital-expertise.nl

Nhien-An Le-Khac,
School of Computer Science
University College Dublin,
Ireland
an.lekhac@ucd.ie

Thanh Thoa Pham Thi
Computer Science Department
Maynooth University,
Ireland
thoa.pham@nuim.ie

*Abstract*—Today a lot of digital evidences for crime investigation includes a geospatial component. This data comes from various sources such as smartphones, tablets, navigation systems, digital camera with global positioning system (GPS), etc. The geospatial data plays a crucial role in crime investigation such as helping to tracking suspects, profiling serial offenders, recognizing trends in criminal activities, just a few. Many techniques and Geographic Information Systems (GIS) tools have been used to extract, analyse and visualise geospatial data. However, in some specific circumstances, the existing tools are not suitable for use as they don't meet investigators' needs. This paper presents a bespoke forensic GIS tool based on specific requirements of the investigators of a law enforcement Department. Firstly the paper discusses some existing forensic GIS tools/environments in practices, and then it presents some investigators requirements and show the unsuitability of the existing tools. The paper continues with the presentation of the system architecture of the new tool and its components. It also introduces various applications and use cases which have been deploying at the Department as an evaluation of the developed tool.

## I. INTRODUCTION

Nowadays, the spatial data within seized data conducted in crime investigations is ubiquitous. This spatial data can make a huge contribution to investigations such as tracking suspects, profiling serial offenders, recognizing trends in criminal activities to address them at an early stage, among other purposes. For example, in the digital forensic related to online child abuse exploitation investigations, investigators can use packages ZiuZ VizX [1] and Blue Bear LACE [2] software packages to select, categorize and analyse relevant materials. These software can filter huge amounts of pictures and export images with location data, but miss the ability to visualise the pictures with location information.. In general, before using and analysing the spatial information a digital forensic investigator must collect the digital data, process it and also be in charge of proper storage so the chain of custody and evidence is ensured. One last step before the crime investigator or analyst can do their work, is the digital forensic investigator should ensure that the processed spatial data can be showed on a map. However, the most correct way to show that spatial information on a map is still a challenge.

Indeed, there are more uses of mapping software and GIS services in law enforcement (LE) agencies. The fact is a wealth of spatial information is not efficiently exploited. For example, in the digital forensic unit of a LE department, every investigation will be registered and saved in one storage location including seized digital devices and the location where the digital devices are captured. Furthermore the location of camera surveillances and its video images are also registered.

When officers from a crime unit are investigating a specific crime and want to know whether there are images from surveillance cameras available around the crime scene (or the location where the digital device are captured), they will check if the vicinity of the crime scene itself or surveillance cameras are visible. By using mapping software/GIS services this information could be made quickly and effectively visible.

Hence, the combined visualization of different types of spatial information, i.e. crime scene and camera surveillances location, can be of great value in investigating criminal activities. From the investigators perspective based on the above example, it is possible to find the creator of the seized photo's when he / she is caught on surveillance cameras at the moment they were taken.

Normally, to make this spatial information visible to crime investigators, in the traditional approach, third-party map application such as Google map, MapInfo, ArcGIS, etc. has to be used. However, this method of processing geospatial information is not flexible and efficient enough in terms of making relationships between different sources of geospatial information, e.g. from camera surveillances and seized images. Moreover, these map software/services are normally designed for general purposes rather than for forensic perspectives. For example, the computer systems with these map software/services are required to connect to the Internet. In real-world application, research data and especially an ongoing investigation, cannot and should not be viewed on a computer connected to the Internet. In addition, we do not know whether a computer connected to the Internet is compromised.

Besides many other desktop GIS tools focus on analysis and statistic, which do not suit in this case.

Therefore, in this paper, we firstly identify requirements of map software/services from the investigator prospective. We then describe a comparative study of different GIS environments based on requirements discussed. We also present the architecture and the GIS forensics tool that addressed the mentioned requirements.

The rest of this paper is organised as follows: Section II discusses background and related work in this area. We point out the forensic requirements of GIS environments and show the unsuitability of the existing tools in Section III. Section IV proposes an in-house GIS environment addressing these requirements. Section V introduces use cases or applications have been currently deploying at the Department, which are considered as the utility evaluation of the tool. Finally, we conclude and discuss future work in Section VI.

## II. BACKGROUND

### A. Definition

First of all, we are looking at the definition of different terms related to the forensic science with spatial data [3]:

- Forensic geography is a sub-discipline of geography wherein a geographer or other expert does research and provides expert testimony appropriate to a court of law based on geography theory and principles [4] [5].
- Geoforensics or Forensic geoscience are synonymous with each other and refer to the application of geography or earth science to forensic investigations [6][7][8][9].
- Forensic mapping is a field of forensic geography that maps criminal activities using location data from GPS devices and cell phone usage data [5].
- Environmental forensics focuses exclusively on environmental concerns and enforcement [10][11].
- Forensic GIS is to provide associative evidence, which assist in either proving or disproving links between people, places and objects as they relate to the court of law. In other words Forensic GIS concerns the application of geographic and spatial tools, principles and methodologies to investigate and establish facts within the boundaries of forensics [3].

The latter described discipline, forensic GIS, covers the subject of this paper best, although the boundaries of forensic GIS are not entirely clear, and also has overlap with the other mentioned disciplines. Digital forensics and forensic GIS have great similarities with each other. The spatial data in criminal investigations comes almost always from digital sources. In fact, using spatial technology in digital forensics means to apply spatial technology in investigating and establishing facts that can be presented in a court of law. Elmes et al. [3] describe that there are six phases within forensic GIS. The collection phase of digital forensics is divided into two phases in forensic GIS, namely: Capture and Manage. Also, the examination phase is split into two phases, namely Integration and Manipulation phase. The fifth and sixth phases in forensic GIS are respectively the Analysis and Display spatial data phase, which correspond to the analysis and reporting phase in digital forensics.

### B. Related work

A lot of GIS tools for spatial data forensics have been developed and used in practices and academia. They can be classified in four categories: i) tools for statistic and analysis of spatial data in forensics without geographic mapping ii) tools mainly for statistic and analysis of spatial data in forensics with little geographic mapping enabling iii) tools for mainly crime mapping with little analysis or manual analysis by investigators based on the visual communication and iv) tools fully combining the analysis and crime mapping. Table 1 classifies some existing tools from our study.

Table 1. Existing forensic GIS tools/environments characteristic

| GIS tools/apps | Analysis and Statistic function | Mapping Enabling |
|---|---|---|
| Crimelinkage | ● | ○ |
| CrimeStat | ● | ○ |
| Urban Crime Simulator | ● | ○ |
| SPIDER Crimes Series Analysis | ● | ○ |
| GeoDa | ● | ◐ |
| Web app ref 1 | ○ | ● |
| Web app ref 2 | ○ | ● |
| ArcGIS | ● | ● |
| MassStats/Maptitude | ● | ● |

Legend:

● *fully implemented*   ○ *unimplemented*   ◐ *partially implemented*

Some representatives of the first category are as follows:

*Crimelinkage [12]*: This package specifically includes methods for criminal case linkage, crime series identification and clustering, and suspect identification.

*CrimeStat [13]:* CrimeStat is a spatial statistical program that is used to analyse the locations of crime incidents and identify hot spots of spatial objects in a GIS – as it can interface with GIS.

*Urban Crime Simulator [14]*: Urban Crime Simulator groups urban neighbourhoods into clusters based on a user's definition of neighbourhood characteristics. The software then uses the resulting clusters to estimate changes in crime rates.

*SPIDER Crime Series Analysis Software [15]*: SPIDER is a spatial statistics program for the tactical crime analysis of linked crime incident locations. The purpose of SPIDER is to provide spatial and temporal analysis support to law enforcement agencies investigating a series of linked crimes. On the results of the analysis investigative decision could be made.

*GeoDa [16]* represents the second category, GeoDa is a collection of software tools designed to implement techniques for exploratory spatial data analysis on lattice data. It also provides crime hotspot mapping functions and it is a stand alone application.

The third category focuses on crime mapping and less analysis, or investigators carry out manual analysis. For instance, [17] allows to render crime locations on maps using four hotspot mapping techniques which are choropleth mapping, grid mapping, spatial ellipse mapping and kernel density mapping using Web-based application. The online Crime mapping and Statistic application of Police in San Diego [18] is opened to public for data communication about crime statistics, hotspot overall the city.

The last category completed the two aspects is represented by the commercial ArcGIS package and Maptitude/Masstats [19]. ArcGIS is a commercial fully featured geospatial software package. With ArcGis maps are included and can be maintained on a separate server. Masstats is also a commercial full functions GIS for law enforcement to mapping and analysing data.

The majority of the listed software packages are geospatial analysis, statistics, or profiling tools. Which do not provide map data where spatial data can be plotted.

### III. REQUIREMENTS FOR A FORENSIC GIS ENVIRONMENT

The benefits of rendering spatial data on maps are huge, in particular for forensic investigators. Maps services can be deployed on stand-alone or desktop applications as well as web-based applications.

A web-based application is more convenient to distribute over the law enforcement unit over the Internet protocol. However, because of the nature of the task which requires high safety, privacy and confidential, the application should not connect to the Internet but rather the Intranet. Therefore there is a need for web-based mapping application on the Intranet.

In addition, the seized spatial data component from the evidences may come from different sources and come in various formats. Based on such specific context, such an Intranet application must satisfy the following requirements:

R1. The application/tool cannot connect to the Internet when the investigator is working to prevent any malicious attack. Which means the Web Map Services are out of choice and there is a need to develop the own maps service.

R2. The developed map services must be available for an Intranet environment to use in the law enforcement department.

R3. The tool should allow displaying different spatial data format such as GML, KML, GPX on maps because of the variety of spatial data sources.

R4. As a consequence, the tool allows integration of various spatial data format and display on maps.

R5. It has the ability to import open geodata sets as now there are lot of open geo data available which is useful for investigation

R6. It should be able to get data from other sources such as wifi-scans, camera surveillances, etc.

R7. It should be an open source solution to reduce cost.

Based on these requirements, an evaluation of the existing tools which have mapping functions is carried out as presented in Table 2. The tools without mapping elements are out of our evaluation as they completely do not suit.

The first two software packages do not meet six of the seven identified requirements. The only requirement to which they meet is that they are free to use. ArcGIS and Maptitude do have maps and satisfy the requirements but they are not free. ArcGIS also provides different types of geospatial file formats which are created in sight corpse. And self-generated geo information can also be visualized on the maps.

Table 2. GIS enviroments vs. requirements.

| GIS Environment | R1 | R2 | R3 | R4 | R5 | R6 | R7 |
|---|---|---|---|---|---|---|---|
| GeoDa | + | - | + | - | - | - | + |
| Web-based GIS [17] | - | + | + | - | - | - | + |
| ArcGIS | + | + | + | + | + | + | - |
| MassStats/Maptitude | + | + | + | + | + | + | - |

Table 2 shows that there is no suitable tools for us to use, it is needed to develop an in-house system.

The requirement that no connection to the Internet may be made to the LE network. The only solution that's free of charge is the one with an OpenStreetMaps web map services solution.

### IV. PROPOSED FORENSICS GIS ENVIRONMENT

The proposed system needs a dedicated Maps Service which act as a map server. The server contains tile maps, which are extracted from OpenStreetMaps. By this way, there is no need to access to the Internet. The Map service is made available for the Intranet environment. A virtual machine on the server is also needed to provide GIS services and back-end processes. The front-end side deals with web-based interfaces where the investigators can do their work with the system. The system architecture is drawn in Figure 1.

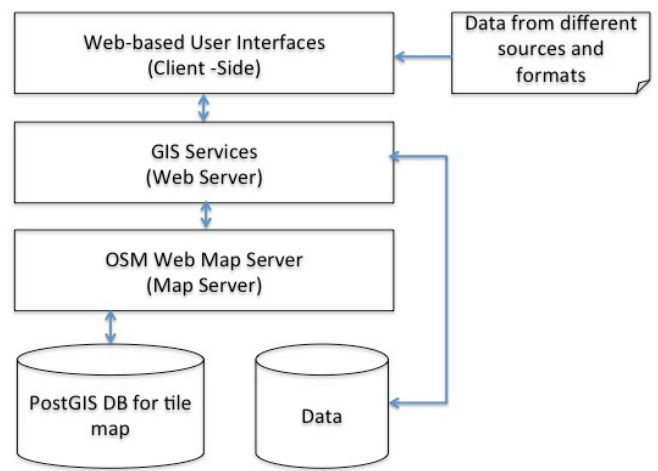

Fig. 1. GIS environment framework

## A. Building the tile server

As mentioned, we chose a tile server based on OpenStreetMap (OSM) as a basis for the forensic GIS. The main components of the OpenStreetMap tile server are:

- Mod_tile is a system to serve raster tiles. It provides a dynamic combination of efficient caching and on the fly rendering. Due to its dynamic rendering, only a small fraction of overall tiles need to be kept on disk, reducing the resources required.
- Renderd is a daemon process to serve tile requests in a queue.
- Mapnik is a tool kit for rendering maps. It is designed to be fast and is suitable for tile generation on high-end servers.
- osm2pgsql is used to convert OpenStreetMap data to postGIS-enabled PostgreSQL databases. To store raster data/map tile.
- postgresql/postgis is used as the underlying database

## B. Building GIS Services

The GIS services deal with back-end services and send data to client side. GIS services and tile server are separated from each other. The tile server is physically housed at the data centre. The GIS component of the framework will be implemented within a virtual machine. These services can be easily accessed among digital specialists from LE departments. These services can be maintained and upgraded easily.

## C. The GIS front-end:

The GIS front-end is web-based. The found spatial data can be imported from an html page and then the spatial data will be displayed on the map. For making the spatial data visible in layers on the map some special spatial libraries must be used such as OpenLayers and Leaflet. We choose Leaflet because of lightness of the code, the flexibility and efficiency. However, it should be noted that Leaflet can easily be replaced without many modifications if Open Layers at a later time is found to be better applicable.

## V. USE CASES AND APPLICATIONS

The developed forensic GIS system has been currently deploying in the Department. This section presents different use cases with that system.

### A. Cameras near crime scene

The first application relates to a cameras near crime scene detection. This application brings all (private) cameras overlooking public spaces mapped on an interactive map. Because of privacy regulations only the police officers can see that data. For example, when there has been a robbery, mugging or a burglary, the police officers can use the interactive map to quickly see if there is camera footage of the incident. If so, the police officer will contact the owner of the camera. Before developing this system, the functionality is based on Google Maps in which we can "pin" camera setups. These are public and private cameras that have a view of the open road. The system offers especially benefits after of detection an incident, however the information regarding private and public cameras are still zero.

On the other hand, the business registration system of the police department with the database of "Camera Picture" is filled with operational information, a rich source of information regarding the existing cameras. Using the proposed framework, in the representation of cameras at a crime scene in a layer to the OpenStreetMap server may indicate where the crime scene is standing with one mouse click. A radius for the crime scene can be given in meters. Any known camera within the specified radius is then displayed on-map. It is also possible not to display cameras in a particular category. Figure 2 shows that in the vicinity of the Central Station in The Hague, public and private cameras are present.

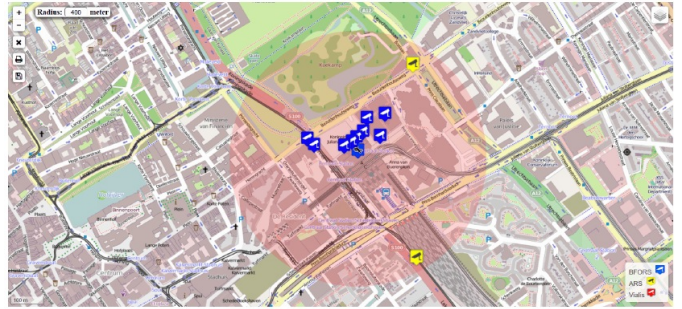

Fig. 2. Camera's nearby a crime scene.

In addition all known information about the cameras is stored and it is displayed when clicking on the blue icon.

### B. Visualizing Wi-Fi scans & Searching (B)SSIDs

Wireless Fidelity, abbreviated Wi-Fi is a technology which widespread in today's society. Almost every household or company does have one or more wireless access points. The modems delivered by Internet Service Providers (ISP) are standard equipped with Wi-Fi functionality.

To connect to a Wi-Fi access point, it sends out its name, called the Service Set Identifier (SSID). It is an identification method used in the IEEE 802.11 (Wi-Fi) standard. SSID makes it possible to separate wireless computer networks from each other, to give each through a separate wireless network name (SSID).

With specific software, like Kismet or Airodump-ng, SSID and other interesting data about the wireless network can be captured. By making use of a GPS device with a Wi-Fi scan not only the SSID, but also the BSSID, the MAC address of the access point and the GPS location of the access point can be found (Figure 5). A seized digital device with Wi-Fi functionality of a suspect can be examined of this device have been connected with any Wi-Fi network. By comparing the results of the post mortem investigation with results of a Wi-Fi scan there can be determined of a suspect was on a location on a certain day, date and time.

Also, different Wi-Fi scans are compared with each other. Are there new wireless networks come in, there are wireless networks gone or are there who have been given a different name? This can also be made visual on a map with pin of different color, as shown in Figure 3.

When scanning for wireless networks it may be possible that BSSIDs are collected from mobile devices with Wi-Fi functionality, that are associated or not associated with wireless networks. This can be helpful if obtained and stored BSSIDs can be searched on a Wigle [20] like way. Searching for BSSIDs (MAC addresses) from those mobile devices can help in crime investigations.

As an example, during a burglary in a home several computers were stolen. One computer was an Apple MacBook Pro and the MAC address was known. After performing some Wi-Fi scans preformed in a particular neighbourhood, the MAC address of the Apple MacBook Pro was found. Subsequently, as a result, the burglar has been arrested and several burglaries were solved.

### C. ANPR and Bluetooth

Many mobile devices, such as phones, navigation systems, tablets, car kits, earphones and entertainment systems in cars, are equipped with Bluetooth. Bluetooth is continuously sending out identification signals. Research shows that one in 42% of the passing vehicles or Bluetooth systems are detectable [21]. The "Verkeers Informatie Dienst (VID)"10 is the traffic specialist of the Netherlands. The VID measures the traffic, manufactures and delivers traffic information to governments, events, media and other business customers who need traffic information. The VID has an advanced, nationwide monitoring network for providing insight into the traffic. The VID uses proprietary, so-called "Bluetooth Meetsysteem (VBM)"11 to measure the traffic information, travel times and route choice behavior. With this method 30 to 35% of the traffic can be followed. When measuring various data is stored, including the MAC address of the Bluetooth device.

This system of the VID can also be used as a good tool in tracking suspects and / or missing persons. A MAC address of the Bluetooth device is in the first instance not directly identifying. Or it must be known is searched for which MAC address, and by whom the MAC address is in use, for example, it is known what the Bluetooth MAC address is of the mobile phone of a person that is wanted.

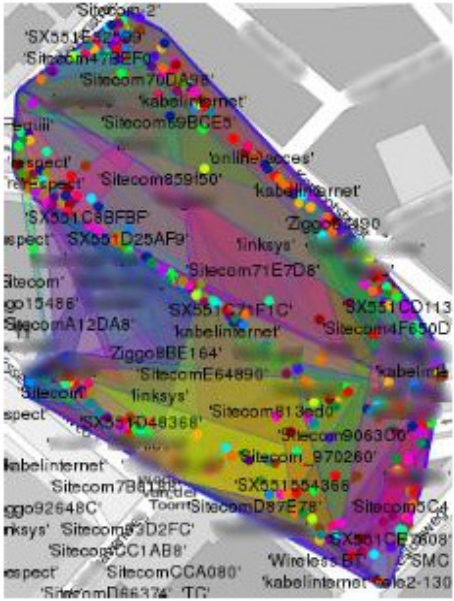

Fig. 3. Visualization of a WiFi scan.

Also, different Wi-Fi scans are compared with each other. Are there new wireless networks come in, there are wireless networks gone or are there who have been given a different name? This can also be made visual on a map with pin of different color, as shown in Figure 4.

But combined with the Automatic number plate recognition (ANPR) system that is in place by many A and B roads located in the Netherlands for route controls by the authorities, the identification of an individual person comes very close. With the combination of the location information and the corresponding times of registration, a Bluetooth MAC address can be associated with a license plate. Which is obviously to a great service to law enforcement.

The spatial insight from the combination of information from the Bluetooth measurement system and Automatic Number Plate Registration system gives crime investigators greater clarity in solving crimes.

### D. GPS track with timeline

In police investigations occasionally GPS trackers are used, where it can be tracked vehicles. If the collected GPS data is visualized on a map in combination with timeline functionality, quickly distinctly places where a vehicle has stopped and for how long. The most interesting locations for further investigation are so identified. See Figure 5 for a brief overview.

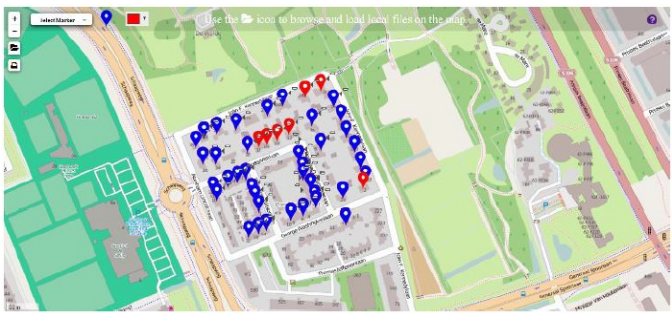

Fig. 4. Comparing Wi-Fi scans.

When scanning for wireless networks it may be possible that BSSIDs are collected from mobile devices with Wi-Fi functionality, that are associated or not associated with wireless networks. This can be helpful if obtained and stored BSSIDs

## VI. Conclusion and Future Work

In this paper we review different GIS environments for forensics analysis and mapping. Even though a number of tools/environments have been in use in practice, there is still a need to develop bespoke system/environment to tailor to investigators needs. As an example seven requirements from investigators have been identified that no tools can meet.

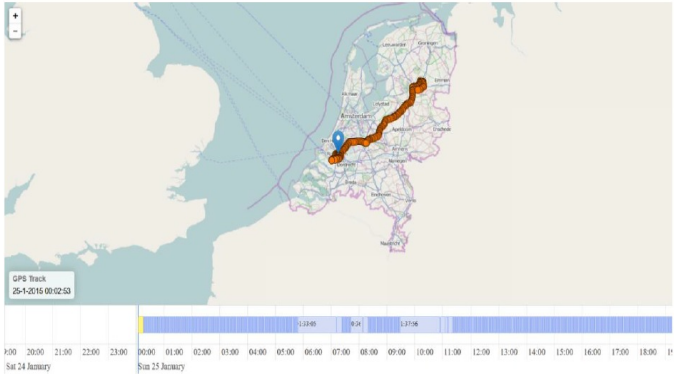

Fig. 5. GPS track with timeline.

We have presented the architecture and its components of a new tool, which has been using in a law enforcement department. The tool satisfies all the seven requirements. With that system the digital investigators, who are generally good at visual thinking, can now bring their thoughts to other crime investigators.

In the future, the system should be deployed as integrated package software with different functions and application types for more convenient uses. In an attempt to combine them into one package, there is a need of consistency consideration in the GUI design and the conformity of the GUI across applications and functionalities.

The developed system is also a good practice to share to the community who may have the same context of work.